\documentclass[conference]{IEEEtran}
\IEEEoverridecommandlockouts
\usepackage{cite}
\usepackage{amsmath,amssymb,amsfonts}
\usepackage{algorithmic}
\usepackage{graphicx}
\usepackage{textcomp}
\usepackage{subcaption}
\usepackage{xcolor}
\usepackage{booktabs}
\def\BibTeX{{\rm B\kern-.05em{\sc i\kern-.025em b}\kern-.08em
    T\kern-.1667em\lower.7ex\hbox{E}\kern-.125emX}}
\begin{document}
\def\name{UEPPR}
\def\fullname{Unified Embedding Based Personalized Product Retrieval}
\title{Unified Embedding Based Personalized Retrieval in Etsy Search\\
}

\author{\IEEEauthorblockN{Rishikesh Jha\textsuperscript{\dag}}
\IEEEauthorblockA{
\textit{Etsy Inc.}\\
Brooklyn, USA \\
rjha@etsy.com}
\and
\IEEEauthorblockN{Siddharth Subramaniyam\textsuperscript{\dag}\textsuperscript{*}}
\IEEEauthorblockA{
\textit{Etsy Inc.}\\
Brooklyn, USA \\
ssubramaniyam@etsy.com}
\and
\IEEEauthorblockN{Ethan Benjamin\textsuperscript{\dag}}
\IEEEauthorblockA{
\textit{Etsy Inc.}\\
Brooklyn, USA \\
ebenjamin@etsy.com}
\and
\IEEEauthorblockN{Thrivikrama Taula}
\IEEEauthorblockA{
\textit{Etsy Inc.}\\
Brooklyn, USA \\
ttaula@etsy.com}
\thanks{\textsuperscript{\dag}Authors contributed equally to this work. \textsuperscript{*}Corresponding Author}
}

\maketitle

\begin{abstract}
Embedding-based neural retrieval is a prevalent approach to addressing the semantic gap problem, which often arises in product searches with tail queries. In contrast, popular queries typically lack context and have a broad intent where additional context from users' historical interaction can be helpful. In this paper, we share our novel approach to address both: semantic gap problem with an end-to-end trained model for personalized semantic retrieval. We propose learning a unified embedding model incorporating graph, transformer and term-based embeddings end-to-end and share our design choices for optimal tradeoff between performance and efficiency. We share our learnings in feature engineering, hard negative sampling strategy, and application of transformer model, including a novel pre-training strategy and other tricks for improving search relevance and deploying such a model at industry scale. Our personalized retrieval model significantly improves the overall search experience, as measured by a 5.58\% increase in search purchase rate and a 2.63\% increase in site-wide conversion rate, aggregated across multiple A/B tests - on live traffic.
\end{abstract}

\begin{IEEEkeywords}
product search, semantic search, personalization, information retrieval
\end{IEEEkeywords}

\section{Introduction}
Search is a critical feature for users on Etsy, enabling them to find products of interest among more than 100 million vintage, handmade and craft items. At a high level, product search at Etsy is powered by an initial candidate retrieval stage which selects several thousand items from our catalog based on the search query and user context, and subsequent ranking stages which filter and re-rank items from previous steps until a final result set is produced.
Historically, candidate retrieval has been powered by lexical matching engines, which were state of the art prior to recent advances in neural information retrieval and are still an important component of present day search systems\cite{clear}. Lexical matching implementations are highly efficient and scalable, but are known to suffer from a vocabulary gap problem\cite{niramazon} (i.e, product vocab does not match search-query vocab). Additionally, it is difficult for lexical retrieval to account for diverse user-specific preferences\cite{nirtaobao}.
\begin{figure}
    \centering
    \begin{subfigure}{.49\textwidth}
        \centering
        \includegraphics[width=1\linewidth]{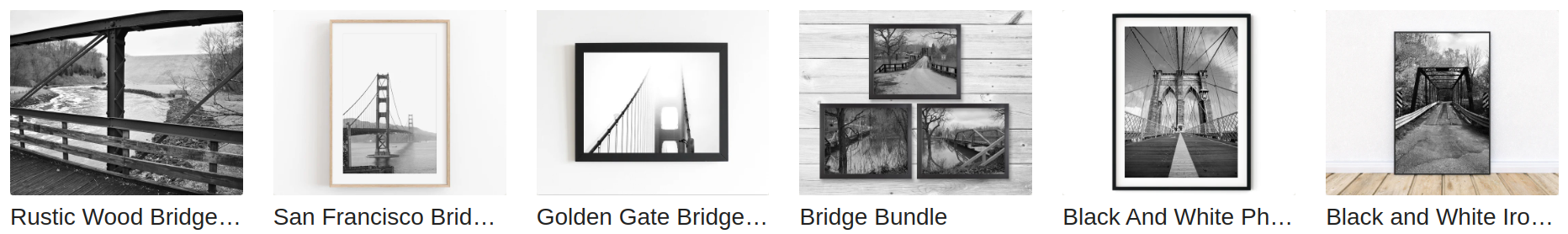}
        \caption{Query without any personalization features}
        \label{fig:sub1}
    \end{subfigure}\\     
    \begin{subfigure}{.49\textwidth}
        \setcounter{subfigure}{1} 
        \centering
        \includegraphics[width=1\linewidth]{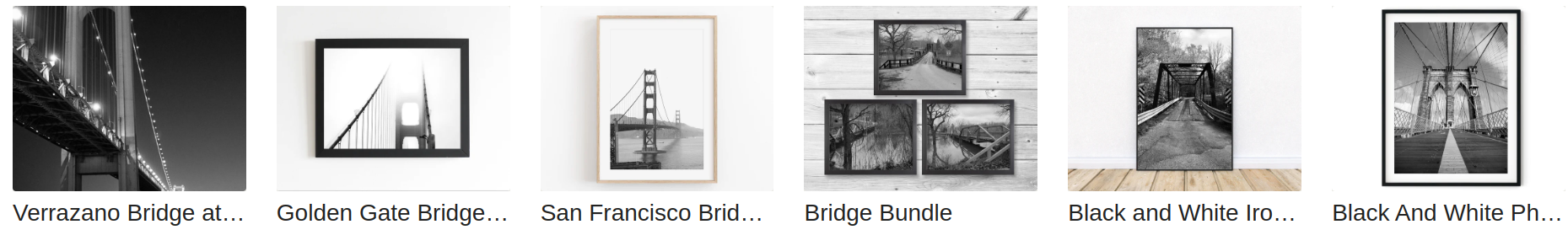}
        \caption{Query with NYC location features without any past queries}
        \label{fig:sub2}
    \end{subfigure}\\
    \bigskip
    \begin{subfigure}{.49\textwidth}
        \setcounter{subfigure}{2}
        \centering
        \includegraphics[width=1\linewidth]{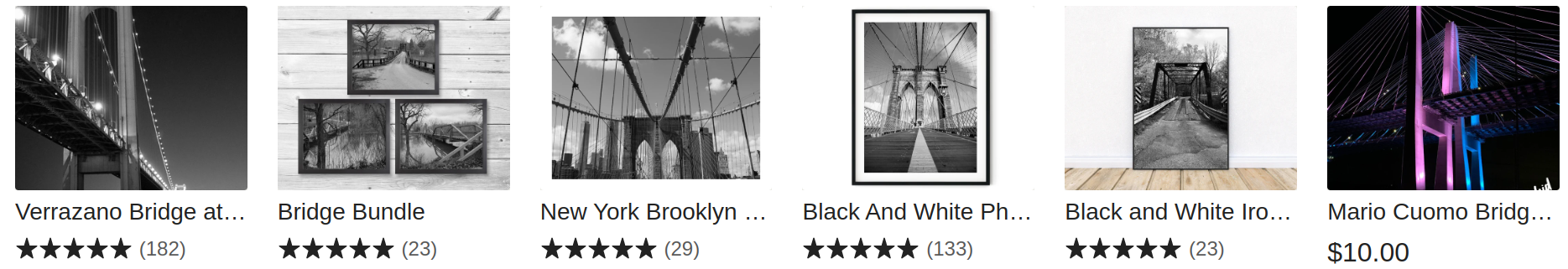}
          \setlength{\abovecaptionskip}{-5pt}
        \setlength{\belowcaptionskip}{-2pt}
        \caption{Query with Brooklyn location features and recent queries include "new york"}
        \label{fig:sub3}
    \end{subfigure}\\
    \setlength{\belowcaptionskip}{-15pt}
    \caption{ Comparing \name's top results for 3 different search sessions demonstrating gradual increase in personalization for the query "bridge photos". }\label{fig:main}
\end{figure}
Recently, neural embedding based retrieval systems have addressed the vocabulary gap problem in web and product search successfully \cite{niramazon,nirfacebook,nirinstacart,nirwalmart}. Neural Retrieval is typically formulated as a text based retrieval problem where a two-tower model (such as \cite{nirdssm}) is used to encode query and document independently. While transformer based language models are suitable for such text retrieval tasks \cite{Devlin2019BERTPO,t5raffel2020exploring} they aren't suited for i) low latency real time inference and ii) short queries with limited context typically found on e-commerce sites. In addition, product search needs to consider other important aspects of a product than text such as its price, seller, location, quality etc. Further, models incorporating user context \cite{CroftTEM,CroftZAM} have also been successfully deployed in production retrieval systems for improving overall search experience \cite{jd.com,nirtaobao} Particularly, in-session user history can be useful context for understanding broad intent search queries though they also require a low latency model for real time inference. In addition, graph based embedding learning approaches specifically on bi-partite graphs of query and product provide complementary information and improve product representations by leveraging neighbor query information \cite{graphsage,NGCF,gcnamazonlu2021graphbased}. Specifically for queries like "mothers day gift" which are common on Etsy, retrieving relevant products just based on text features can be difficult. Thus, we propose learning a unified embedding model incorporating graph, transformer and term-based embeddings end-to-end for leveraging the best of all these approaches and addressing the dual problem of vocabulary gap and retrieving personalized products at scale.

Another consideration beyond textual similarity in e-commerce search systems is inherent product quality independent of query (e.g average product review)\cite{ecomranking}. However, prior work in embedding-based retrieval eschews explicit modeling of product quality, leaving this responsibility entirely to later stages. Oftentimes, even with additional context from specific users there can be more semantically relevant products in the catalog than what can be efficiently re-ranked by subsequent layers. Thus, to produce an optimal set of retrieval candidates, it is beneficial to consider quality as long as relevance is not sacrificed. To this end, we propose \textit{ANN-Based Product Boosting} where we augment existing embeddings with vectors tuned to balance quality and relevance when combined with inner product scoring. We employ black box optimization to identify globally optimal quality weights which maximize desired target metrics with realistic serving constraints.

In this paper, we present the details of training and deploying our {\fullname}(\name) system for retrieving personalized products. We demonstrate the efficacy of such a system on both head and tail queries with both offline evaluations and live A/B testing. 
Our contributions can be summarized as follows:

i) We present our novel two-tower model with a unified embedding based product encoder and joint user-query encoder deployed on an e-commerce website to handle diverse range of search queries.

ii) We share our novel training strategies including mining hard negatives during training, our pre-training strategy for language models used for product text feature representation and other design choices such as loss function and approach to generating training data for training such a model.

iii) Using ablation studies, we share our learnings in using diverse feature encoders and engineering tricks for encoding a variety of user, product and query features.

iv) Our novel approach of \textit{ANN-Based Product Boosting} for re-ranking candidates based on listing quality scores.
\section{RELATED WORK}

In recent years, deep learning techniques have been widely applied to information retrieval (IR) systems, specifically in neural-augmented lexical retrieval. These techniques can be categorized into representation models (such as DSSM \cite{nirdssm} and CLSM \cite{clsm}) and query-document interaction models (such as MatchPyramid \cite{matchpyramid} and Match-SRNN \cite{matchsrnn}).

Most e-commerce search systems employ the two-tower architecture with an ANN for embedding based retrieval (EBR) \cite{niramazon,nirfacebook,nirwalmart,nirinstacart,nirtaobao}]. Nigam, Song et al\cite{niramazon} leveraged shallow textual bag of word embeddings and categorical product feature embeddings. Recent work has sought to improve the representational capacity of EBR models \cite{CroftZAM,CroftTEM}. Personalized models incorporate user behavior and profile features to better tailor candidates for specific users \cite{jd.com,nirtaobao}. Models enriched by graph-based embeddings \cite{graphsage,NGCF} improve on purely content-based models by leveraging known relationships between distinct entities\cite{gcnamazonlu2021graphbased,geps}. Thus, we use bipartite graph encodings as a part of our unified representation.

Pre-trained transformer-based language models \cite{Devlin2019BERTPO,zhuang-etal-2021-robustly,xlnetjiang-etal-2020-cross} improve upon shallow unordered term embeddings by enabling contextualized term representations \cite{tk} and deeper semantic understanding (though, increased modeling capacity comes at the cost of latency\cite{xml}, and enriched language understanding may provide less benefit to e-commerce queries which are short and lack context.

\textit{Hard Negative Sampling} is an area of active research focused on improving the training of dense retrieval models by incorporating hard negatives. \cite{hnsdenseeret} proposed an approach that uses hard negative samples to optimize the training of dense retrieval models. Similarly, \cite{hnsnce} proposed an approximate nearest neighbor negative contrastive learning for dense text retrieval. These methods have shown significant improvements in retrieval performance. Facebook in their EBR model\cite{nirfacebook} introduced tricks for hard-negatives mining and benefits of combining them with random negatives.

In summary, deep learning techniques have shown great potential in improving the accuracy and efficiency of IR systems. The above-mentioned works demonstrate the effectiveness of various deep learning techniques in addressing different challenges in IR, including modeling semantic relationships, personalization, and improving the training of retrieval models.

\begin{figure}[h]
  \centering
  \includegraphics[width=\linewidth]{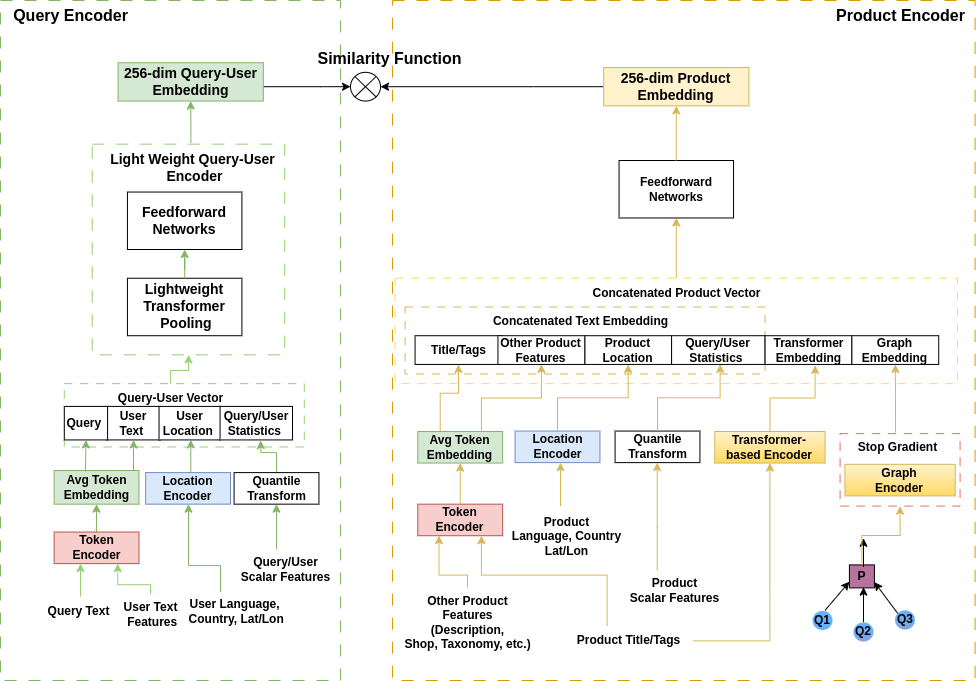}
  \setlength{\abovecaptionskip}{-2pt}
  \setlength{\belowcaptionskip}{-2pt}
  \caption{Base architecture of \name}
  \label{model}
\end{figure}

\section{UNIFIED EMBEDDING MODEL}
\label{uem}
Our proposed two-tower {\name} model consists of a product encoder $\mathcal{P}$ and a joint query-user encoder $\mathcal{Q}$. For a given query $q$, user $u$ we compute the personalized semantic score $y$ for a given product $p$ as follows: \begin{equation}
\label{similarity} y = cos(\mathcal{P}(p),  \mathcal{Q}(q,u))
\end{equation} where both $\mathcal{P}(p)$ and $\mathcal{Q}(q,u)$ are d-dimensional vectors. Our choice to use cosine as the similarity function is based on efficiency since that keep the two-towers independent and allows offline indexing and efficient serving via ANN. Representations from $\mathcal{P}$ and $\mathcal{Q}$ are a result of pooling of constituent encoders used for representing relevant features. 
Our goal is to maximize Recall@K, a metric used widely in literature since it effectively measures retrieval performance. 
In this section, we present the architecture details of our proposed {\name} model, along with the key design choices in feature engineering. We begin by describing the design of our product encoder, followed by an explanation of the shared encoders between both towers. Subsequently, we provide a comprehensive overview of the joint query-user encoder.

Furthermore, we delve into the critical approaches that were instrumental in the successful offline training of $\mathcal{P}$ and $\mathcal{Q}$. These include our novel loss function, the pre-training strategy employed for language models, the technique for mining positives from search logs to create labeled data, and our method for mining dynamic hard negatives.

\subsection{Product Encoder}
In Figure \ref{model} (shown on the right side), we present the product encoder $\mathcal{P}$, which is designed as a unified embedding model to capture various complementary aspects of a product. The product representations obtained from this encoder are pooled through concatenation and further projected using successive layers of feedforward networks and Batch Normalization to obtain the final product representation. In this section, we focus on describing the encoders dedicated to encoding crucial product features, as well as the relevant feature engineering techniques employed.

\subsubsection{Transformer based representations}
Contextual representations from transformers can be helpful in biasing product representations towards important tokens from text features like title and tags. However, our findings indicate that fine-tuning two widely used text encoders, distilBERT \cite{distilbertsanh2020} and the encoder part of T5 \cite{t5raffel2020exploring}, did not yield any offline improvements. This could be attributed to our asymmetric architecture, where the query tower lacks an identical transformer due to latency considerations. Following that, we employed a pre-training approach influenced by docT5query \cite{doct5queryCheriton2019FromDT}. In this strategy, we trained a T5-small model to generate the most historically purchased query based on product text and utilized its encoder. Our hypothesis is that this approach would enhance the model's comprehension of crucial aspects such as product title/tag.

\subsubsection{Bipartite graph encoder}
We employ bipartite graph-based embeddings \cite{gcnamazonlu2021graphbased} to effectively address the semantic gap between query terms and product content that is relevant based on historical search. To construct a product-query graph, we utilize over a year of search logs containing positive interactions between queries and products. During training, we sample queries for a given product from the graph and encode them using parameters shared with the query tower. We use average pooling over query embeddings to get a one-hop graph embedding. To ensure generalization, we exclude the target query from the set of sampled neighbors. We also found that sharing parameters led to over-fitting, which we addressed with several interventions and found that simply freezing shared parameter updates due to graph encoder led to best performance.


\subsection{Shared Encoders}
\subsubsection{Token and ID Embeddings}
\label{token_embeddings}
Queries and products each contain sets of token fields $F_q$ and $F_p$. Each token field $f$ consists of a bag of unordered tokens derived from raw inputs. As in \cite{niramazon} we extract unigrams, bigrams, and character trigrams from textual fields. Additionally, we extract product category at each level of hierarchy (e.g \#category\_furniture, \#category\_furniture.bedroom, etc). For product attributes like color or material, we extract tokens for each key/value pair (e.g \#attr\_color\_red), and also extract textual n-grams for attributes where sellers are able to input arbitrary values. 

Our product and query encoders both employ average embedding layers to encode a set of fields $F$ containing $N_t(F)$ total tokens via a simple average of token embeddings:

\begin{equation}
\text{avg}(F) = \frac{1}{N_t(F)} \sum_{f \in F} \sum_{t \in f} E[t]
\end{equation}

Initially, we employed an average embedding approach for the token input layer of our product and query-user towers, represented as $\text{token\_rep}(p) = \text{avg}(F_p)$. However, we observed that incorporating certain new fields, especially noisy ones like keywords extracted from the product description, could lead to a degradation in model performance. To address this, we partitioned the fields into distinct groups, denoted as $F_p^1, F_p^2,...$ (e.g., one group exclusively containing description tokens). We then concatenated the average embeddings of each group, resulting in the updated token representation $\text{token\_rep}(p) = \text{concat}([\text{avg}(F_p^1); \text{avg}(F_p^2); ...])$. 
\subsubsection{Location Encoder}
As in \cite{HuangFacebook}, we incorporate both user location and listing location features in the query and listing towers respectively. We embed language, country, and zip code, and additionally account for location at multiple levels of granularity using both latitude/longitude and zip codes. We perform K-means clustering on customer latitude/longitude coordinates and extract location bucket IDs for users and listings with different values of k (50,100,500). We also capture coarse geographic regions by extracting prefixes of zip codes (e.g ``54321" yields ``5", ``54", etc), as prefixes correspond to contiguous geographic regions. We found that zip codes and location buckets are non-redundant in terms of model performance, possibly because they capture somewhat different notions of proximity (for instance, k-means clusters can have large variability in geographic span due to dependence on population density, while zip codes tend to be more uniform in size). Using all location features together, we observed a relative gain of 8\% in purchase recall for US users.
\subsection{Joint Query-User Encoder}
We illustrate the architecture of our joint tower on the left side in Figure \ref{model}. To represent the query text we use the lightweight token encoder as described in section \ref{token_embeddings}. At the token level these parameters are shared between product title/tags and the graph encoder. In addition, we use location encoders described earlier to represent user location. Further, we incorporate user engagement history from a variety of event types similar to \cite{jd.com} and \cite{nirtaobao}. In particular, we leverage a user's recent searches, recent shop clicks, terms from recently clicked items, and tags of all-time purchased items. To minimize latency on the query-user tower we use lightweight token embedding as described in \ref{token_embeddings} and attention mechanisms to incorporate these in the final joint-query. As pointed out in in \cite{CroftZAM}, not all historical user actions are equally relevant to the current search session, so it can be beneficial to weight events in relation to the current query and to control the total contribution of user events in personalization. Similar to \cite{CroftTEM}, our query-user tower uses lightweight transformers applied at the event level to produce historical embeddings from the query and user actions. In particular, we apply a transformer to the sequence $[q, u^s_1,...,u^s_m, u^c_1 ... u^c_n]$, where $q$ is search query, $u_s$ is a sequence of a user's recent search queries and $u^c$ is a user's recently clicked shops. Each vector in $u_s$ is a simple average of a query's token embeddings, and each vector $u_c$ is the shop ID embedding. We find that a 1-layer transformer contributes little to latency, and improves offline recall by 2\%.




\subsection{Negative Mining} Our negative mining strategy employs three different sources of negative products:
\begin{itemize}
    \item \textit{Hard in-batch negatives}: In-batch negatives are positive products from other queries in a batch that are sampled as negative for a given query. Since product and query representations are independent in a two-tower model they don't need to be recomputed as in a cross encoder. Further, we explored allowing the model to focus on the \textit{hardest} examples by sampling most similar in-batch products for a given query and we found that it works better than uniform sampling.
    \item \textit{Uniform negatives}: In addition, we also sample products uniformly from our entire corpus of products as negatives. Uniform negatives provide complementary value to hard negatives because they help \textit{bottom ranking} performance while hard negatives help top ranking performance \cite{zhan2021optimizing}. We have also observed that size of corpus from which negatives are sampled affects performance and since in-batch negatives only have positive products, diversity in negatives is limited and model can overfit. 
    \item \textit{Dynamic Hard Negatives from large batch}: Dynamic negatives, i.e., negatives with respect to model parameters during training, have been shown to improve dense retrieval performance\cite{zhan2021optimizing} over static negatives. Our memory-efficient approach to sampling dynamic negatives has two steps: First, we randomly sample large batch of negative samples from the product corpus and use current model parameters to select top K most similar products, but do not update model parameters yet. After the initial selection the \textit{dynamic hard negatives} are combined with other sources of negative examples, which are used to update model parameters.
\end{itemize}
Our loss from negative examples is a weighted sum of individual from each mining strategy and we linearly update the weights during training. We warmup training with only uniform negatives and  linearly decay loss weight from uniform negatives and increase weight for hard negatives as we found that it is ideal for convergence and performance.
\subsection{Loss Function}
Threshold-based pruning is a widely adopted approach in the retrieval layer, employed to eliminate irrelevant candidates. Though pairwise loss functions are widely used they aren't suited for threshold based pruning\cite{niramazon}. Our approach incorporates a hinge loss framework to establish a threshold during model training phase itself. Since our training data consists of different kinds of interactions types and each interaction type represents a different degree of relevance, we employ a multi-part hinge loss where each part is associated with a different threshold. Given the output score as $y$ and the true label as $\hat{y}$, our loss function can be expressed as:
\[ L(y, \hat{y}) = \left(\sum_{i \in I} \mathbb{I}[\hat{y} == i] \cdot (-\min(0, y - \epsilon_i))\right) + \mathbb{I-} \cdot max(0, y - \epsilon)\]
where $I$ is set of all positive interactions and $\epsilon_i$ is the threshold corresponding to it and $\mathbb{I-}$, $\epsilon$ are indicator variable and threshold for negative samples respectively.

\section{ANN-BASED PRODUCT BOOSTING}
\label{annproductboost}

When forming a candidate set of products for a query, it is beneficial to retrieve candidates that are both semantically relevant to the query and appealing to customers. Within Etsy, our inverted index candidate retrieval system associates products with a query-independent ``quality score" $Q(p)$, and employs multiplicative boosting to compute a candidate score as the product of it's quality score and query-listing relevance score $R(q,p)$, or $S(q,p) = R(q,p)Q(p)$. This quality score can account for properties such as high product rating, product freshness, and shop conversion rate that are known to increase engagement independently of query-listing relevance.

We implement additive boosting within ANN-based semantic retrieval by enriching our model-derived product vectors with additional numerical features and add corresponding feature weights to query vectors. Given original product embedding $p$ and query embedding $q$, we create hydrated vectors $p' = \text{concat}([p; f(p)])$ and $q' = \text{concat}([q; w])$ where $f(p)$ is a feature vector of numerical features, and $w$ is a constant vector of the same dimension. We then model candidate score as $S(q',p') = \text{dot}(p',q') = \text{dot}(p,q) + \text{dot}(f(p), w)$. For serving, we simply index hydrated product vectors rather than original product embeddings, and query our index with a hydrated query vector.

In principal, we could learn query-side feature weights directly as part of end-to-end model training. However, compared to textual features like query or title text, static quality features like shop popularity are sensitive to negative sampling approaches and can easily over-fit on our proxy metrics without careful tuning and also do not optimize for recall directly. 
Instead we use black box optimization to identify query-side feature weights which directly maximize arbitrary target metrics. In particular, we use skopt\cite{skopt} to perform bayesian optimization to learn query weights which optimize purchase recall on items purchased after our model training window.
Note that query boosting weights are static and not query dependent. In future, we can make boosting weights a function of query.
\begin{figure}[h]
  \centering
  \includegraphics[width=\linewidth]{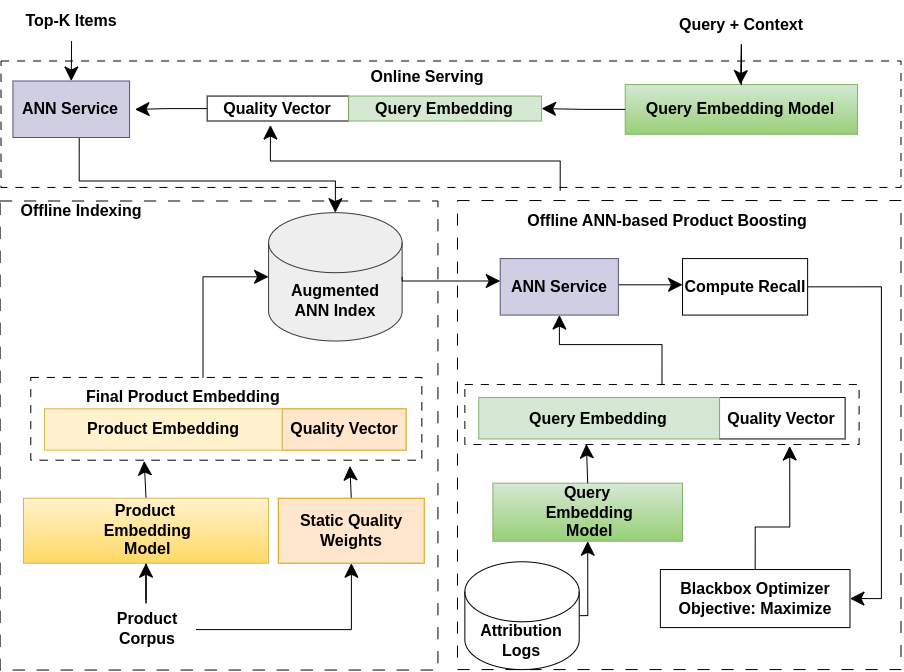}
  \setlength{\abovecaptionskip}{-5pt}
  \setlength{\belowcaptionskip}{-10pt}
  \caption{Online Serving and Offline Indexing of over retrieval service. In addition to the ANN indexing step, we perform ANN-based product boosting as shown on the right}
  \label{serve}
\end{figure}

\section{ONLINE SERVING}
\label{onlineserving}
As part of the retrieval layer, our system should support low-latency (tens of milliseconds) candidate retrieval (query inference plus ANN lookup) for thousands of queries per second. The P99 inference latency for \textit{{\name} boost} is ~18ms. Our approach to addressing challenges related to caching, and ANN tuning that have led to the successful deployment of {\name} is as follows:.
\begin{itemize}
    

\item \textit{Inclusion of in-session features and cache hit rate}. We observed a significant impact on cache hit rate when incorporating in-session features, such as a user's recent queries. To address this, we explored two strategies.i ) \textit{Including ID-level information in the cache key}. By incorporating user/session IDs in the cache key, we maintained a higher cache hit rate, particularly when users paginated or navigated back to the search page. But in order to take advantage of in-session features, this strategy requires the cache TTL to be lowered. ii) \textit{Hashing query context (features) into the cache key}. By performing the feature lookup before cache lookup, we were able to include query context in the cache key. This approach allowed us to maintain a relatively high TTL (cache expiration time) while leveraging in-session features.
    
\item \textit{ANN algorithm}. Using the Faiss library \cite{faissjohnson2017billionscale}, we experimented with various ANN algorithms such as \textit{HNSW} \cite{hnswmalkov2018efficient}, \textit{IVFFlat}, \textit{IVFPQFastScan}, and \textit{IVFPQFastScan} with re-ranking. We measured ANN accuracy using \textit{recall loss@K}, which is the drop in \name's recall metric when going from exact-KNN to ANN. Empirically, we determined that \textit{HNSW} performed well for K values up to 100 (recall loss of 5\%), but this deteriorated as we increased K upto 1000 (recall loss $ > $ 10\%). Modifying the \textit{ef-Construction} \cite{hnswmalkov2018efficient} argument allowed us to trade-off higher memory usage for better recall. Recently, we transitioned to using the 4-bit product quantizer based fastscan index in Faiss, which stores search time look-up tables in registers \cite{fastscancachelocalityandre}. Although this index had lower recall on its own, we achieved improved results by incorporating a re-ranking step after fetching 5 times the required K. We were able to achieve a recall loss of less than 4\% while keeping our P99 latency under 20ms in production.

\item \textit{Tuning ANN parameters}. Following the observations mentioned in \cite{HuangFacebook}, we found it beneficial to tune the ANN system whenever significant changes occurred in the model. We built an offline evaluation pipeline to efficiently tune these parameters using black-box optimization on Google Vizier \cite{viziergolovin2017google}.

\end{itemize}

\section{EXPERIMENTS AND RESULTS}
\begin{table}[h]
  \centering
  \setlength{\abovecaptionskip}{1pt}
  \setlength{\belowcaptionskip}{-4pt}
  \caption{Comparison of aggregate Recall@100 and Recall@1000 on all query segments as well as Recall@100 on Head and Tail query segments.}
  \label{tab:main_result}
  \begin{tabular}{lcccc}
    \toprule
    Method & Recall@100(All) & Head & Tail & Recall@1000(All) \\
    \midrule
    BM25 & 0.55 & 0.63 & 0.45 & 0.62 \\
    Base & 0.33 & 0.32 & 0.42 & 0.61 \\
    Large vocab & 0.45 & 0.45 & 0.49 & 0.72 \\
    Random-walk & 0.6 & \textbf{0.74} & 0.52 & 0.72 \\
    {\name} rand & 0.618 & 0.66 & 0.63 & 0.8 \\
    {\name} no T5 & 0.67 & 0.68 & 0.68 & 0.81 \\
    {\name} & 0.674 & 0.71 & 0.69 & 0.82 \\
    {\name} boost & \textbf{0.708} & 0.72 & \textbf{0.71} & \textbf{0.85} \\
    \bottomrule
  \end{tabular}
  \label{main_result}
\end{table}


    

\begin{table}[h]
  \centering
  \caption{Ablation study showing importance of each negative mining strategy when applied to {\name}}
  \label{tab:recall}
  \begin{tabular}{lcc}
    \toprule
    Method & Recall@100(\%) & Recall@1000(\%) \\
    \midrule
    In-batch[I] + Uniform[U] & - & - \\
    Hard [I] + [U] & 3 & 1.2 \\
    Hard [I] + [U] + Dynamic[D] & 9.3 & 5.1 \\
    Hard [I] + [U] + [D] (weighted) & 11 & 6 \\
    \bottomrule
  \end{tabular}
  \label{tab:hns}
\end{table}
\begin{table}[htbp]
\centering
\caption{Listing Feature Importance in Terms of Relative Improvement in Recall@100 over NIR large vocab model}
\begin{tabular}{lc}
\hline
Listing Feature &  Recall@100 (\%) \\
\hline
Title + Tags (T5)       & 1.00                               \\
Attributes              & 3.90                               \\
Description             & 6.30                               \\
Graph                   & 15.00                              \\
\hline
\end{tabular}
\label{tab:listing_features}
\end{table}
In this section, we present our offline and online evaluation approach and experimental results comparing different design choices and detailed ablation studies and qualitative analysis that aim to provide insights into feature importance and the influence of various feature modeling choices on retrieval performance. We also present our online A/B results showing significant improvements in overall search experience in a large scale e-commerce search-engine.

\subsection{Evaluation Methodology}
\begin{itemize}
    \item \textit{Offline metric} We use Recall@K, a metric used widely in literature since it effectively measures retrieval performance. Essentially, given a query $q$, set of target relevant products $\mathcal{T}$ and set of top K products $\mathcal{R}$: \[Recall@K = \frac{\sum_{t \in \mathcal{T}}\mathbb{I}(t \in \mathcal{R})}{|\mathcal{T}|}\].We have also observed that offline Recall@K metrics positively correlates with online improvement in site-wide and search conversion rates We use $K \in \{100, 1000\}$ for our offline evaluation.
    \item \textit{Training \& Evaluation Data}
    We create our training data by mining positive interactions such as cartadds and purchases from our search logs using past one month of data from across all our platforms. The dataset contains upto 30 millions users and more 150 million interactions. We used TensorFlow2.6 to train our model. We use MirroredStrategy to train model on 4 A100 GPUs and it takes about 72 hours for the model to converge. During evaluation, our objective is to assess the model's capacity to recall purchases in the days following the latest days training data. Consequently, we construct our evaluation dataset by employing sampled search logs that incorporate user purchases occurring over the next day after the training data. Overall, we report average over the next 5 days of evaluation data with each days data consisting of 11k unique queries having purchases. 
    \item \textit{Baseline Model}
    Our baseline embedding-based retrieval model referred as \textit{Base} is a shallow, symmetric two-tower model. This model incorporates term embeddings from raw query text and a product's title and tags. We use a multi-part hinge loss as described in section 4 and select negatives randomly from the entire catalog of listings, as well as from in-batch positives, as in \cite{jd.com}. In addition we also compare our model to highly tuned in-house classical inverted index and random walk based retrievers and demonstrates effectiveness across query segments.
\end{itemize}

\subsection{Offline Experiment Results}
In Table \ref{main_result}, we compare {\name} to strong baselines and with different variants of our own model that show efficacy of individual components. We can see that each of our {\name} outperforms random walk based retrievers and inverted index retrievers on aggregate across query segments while our best non-personalized models are better at tail despite being worse at head queries. Adding personalization to query tower help performance on head queries significantly making it competitive with other methods.\textbf{{\name} rand} refers to our model only uses random and in-batch negatives in training. We can see that purchase recall@100 drops by 7\% without hard negatives. Further, in table \ref{tab:hns} we show ablation showing impact of using three negative mining strategies and show that combining the three hard negative mining strategies with dynamic weighting of loss. \textbf{{\name} boost} refers to model variant that incorporates our proposed ANN-based product boosting. From table \ref{main_result} we can see that boosting leads to overall gains in recall@100 and as hypothesized, it leads to even more significant improvement on head queries than tail, which has more potentially relevant listings.
The \textbf{Large vocab} variant model in table  \ref{main_result} is an improvement over \textbf{Base} in terms of vocab size and more hidden layers. We share an ablation study in Table \ref{tab:listing_features} which shows the importance of each of our individual encoders when added to \textbf{Large vocab} that particularly highlights the significant improvement in performance when adding graph embeddings, as well as the other transformer and term based representations.

\subsubsection{Impact of location embeddings} Our location-aware model tends to produce fewer candidates that are very distant from the user. We observed that the 98th percentile distance to user for candidates in the location enabled model is ~2600 miles, compared to ~4200 miles in the baseline. We observed that for domestic users, our location-based model retrieved significantly fewer international listings. However, the domestic purchase recall limited to international listings was barely affected, suggesting that our model can distinguish exactly which international listings are likely to be relevant and appealing. 

\subsubsection{Qualitative Analysis} In figure \ref{fig:main} we observe the impact of the degree of personalization based on location features and recent searches. For a query "bridge photos", we see that it increase gradually based on user/location features. For a user with location features in NYC/SF, they see more localized results compared to a user with no history. Furthermore, in \ref{fig:main} c, we see that the results are biased even more towards NYC (i.e Mario Cuomo bridge) when the user has recently searched for "NYC" in the session.

\subsection{Online Evaluation}

We evaluated the effectiveness of our system by deploying it online and conducting A/B tests on live Etsy traffic. Our evaluation focused on two important metrics: the relative change in site-wide conversion rate (CVR) and the organic search purchase rate (OSPR). These metrics provide valuable insights into the overall search experience for customers on Etsy. On aggregate, our {\name} system improved CVR by 2.63\% and OSPR by 5.58\%. Notably, the personalized {\name} variants had a greater impact on the signed-in and habitual buyer segments, as expected.

\section{CONCLUSION}
In conclusion, our paper introduces a unified embedding-based personalized product retrieval system, called \name, and provides comprehensive details on its training and deployment. We have successfully demonstrated the system's efficacy through both offline evaluations and live A/B testing, both on head and tail queries.
\bibliographystyle{IEEEtran}
\bibliography{bibliography}

\begin{thebibliography}{10}
\providecommand{\url}[1]{#1}
\csname url@samestyle\endcsname
\providecommand{\newblock}{\relax}
\providecommand{\bibinfo}[2]{#2}
\providecommand{\BIBentrySTDinterwordspacing}{\spaceskip=0pt\relax}
\providecommand{\BIBentryALTinterwordstretchfactor}{4}
\providecommand{\BIBentryALTinterwordspacing}{\spaceskip=\fontdimen2\font plus
\BIBentryALTinterwordstretchfactor\fontdimen3\font minus \fontdimen4\font\relax}
\providecommand{\BIBforeignlanguage}[2]{{%
\expandafter\ifx\csname l@#1\endcsname\relax
\typeout{** WARNING: IEEEtran.bst: No hyphenation pattern has been}%
\typeout{** loaded for the language `#1'. Using the pattern for}%
\typeout{** the default language instead.}%
\else
\language=\csname l@#1\endcsname
\fi
#2}}
\providecommand{\BIBdecl}{\relax}
\BIBdecl

\bibitem{clear}
L.~Gao, Z.~Dai, T.~Chen, Z.~Fan, B.~Van~Durme, and J.~Callan, ``Complement lexical retrieval model with semantic residual embeddings,'' in \emph{Advances in Information Retrieval: 43rd European Conference on IR Research, ECIR 2021, Virtual Event, March 28--April 1, 2021, Proceedings, Part I 43}.\hskip 1em plus 0.5em minus 0.4em\relax Springer, 2021, pp. 146--160.

\bibitem{niramazon}
\BIBentryALTinterwordspacing
P.~Nigam, Y.~Song, V.~Mohan, V.~Lakshman, W.~A. Ding, A.~Shingavi, C.~H. Teo, H.~Gu, and B.~Yin, ``Semantic product search,'' in \emph{Proceedings of the 25th ACM SIGKDD International Conference on Knowledge Discovery \& Data Mining}, ser. KDD '19.\hskip 1em plus 0.5em minus 0.4em\relax New York, NY, USA: Association for Computing Machinery, 2019, p. 2876–2885. [Online]. Available: \url{https://doi.org/10.1145/3292500.3330759}
\BIBentrySTDinterwordspacing

\bibitem{nirtaobao}
\BIBentryALTinterwordspacing
S.~Li, F.~Lv, T.~Jin, G.~Lin, K.~Yang, X.~Zeng, X.-M. Wu, and Q.~Ma, ``Embedding-based product retrieval in taobao search,'' in \emph{Proceedings of the 27th ACM SIGKDD Conference on Knowledge Discovery \& Data Mining}, ser. KDD '21.\hskip 1em plus 0.5em minus 0.4em\relax New York, NY, USA: Association for Computing Machinery, 2021, p. 3181–3189. [Online]. Available: \url{https://doi.org/10.1145/3447548.3467101}
\BIBentrySTDinterwordspacing

\bibitem{nirfacebook}
J.-T. Huang, A.~Sharma, S.~Sun, L.~Xia, D.~Zhang, P.~Pronin, J.~Padmanabhan, G.~Ottaviano, and L.~Yang, ``Embedding-based retrieval in facebook search,'' in \emph{Proceedings of the 26th ACM SIGKDD International Conference on Knowledge Discovery \& Data Mining}, ser. KDD '20.\hskip 1em plus 0.5em minus 0.4em\relax New York, NY, USA: Association for Computing Machinery, 2020, p. 2553–2561.

\bibitem{nirinstacart}
Y.~Xie, T.~Na, X.~Xiao, S.~Manchanda, Y.~Rao, Z.~Xu, G.~Shu, E.~Vasiete, T.~Tenneti, and H.~Wang, ``An embedding-based grocery search model at instacart,'' 2022.

\bibitem{nirwalmart}
A.~Magnani, F.~Liu, S.~Chaidaroon, S.~Yadav, P.~Reddy~Suram, A.~Puthenputhussery, S.~Chen, M.~Xie, A.~Kashi, T.~Lee, and C.~Liao, ``Semantic retrieval at walmart,'' in \emph{Proceedings of the 28th ACM SIGKDD Conference on Knowledge Discovery and Data Mining}, ser. KDD '22.\hskip 1em plus 0.5em minus 0.4em\relax New York, NY, USA: Association for Computing Machinery, 2022, p. 3495–3503.

\bibitem{nirdssm}
\BIBentryALTinterwordspacing
P.-S. Huang, X.~He, J.~Gao, L.~Deng, A.~Acero, and L.~Heck, ``Learning deep structured semantic models for web search using clickthrough data,'' in \emph{Proceedings of the 22nd ACM International Conference on Information \& Knowledge Management}, ser. CIKM '13.\hskip 1em plus 0.5em minus 0.4em\relax New York, NY, USA: Association for Computing Machinery, 2013, p. 2333–2338. [Online]. Available: \url{https://doi.org/10.1145/2505515.2505665}
\BIBentrySTDinterwordspacing

\bibitem{Devlin2019BERTPO}
J.~D. M.-W.~C. Kenton and L.~K. Toutanova, ``Bert: Pre-training of deep bidirectional transformers for language understanding,'' in \emph{Proceedings of naacL-HLT}, vol.~1, 2019, p.~2.

\bibitem{t5raffel2020exploring}
C.~Raffel, N.~Shazeer, A.~Roberts, K.~Lee, S.~Narang, M.~Matena, Y.~Zhou, W.~Li, and P.~J. Liu, ``Exploring the limits of transfer learning with a unified text-to-text transformer,'' \emph{J. Mach. Learn. Res.}, vol.~21, no.~1, jan 2020.

\bibitem{CroftTEM}
K.~Bi, Q.~Ai, and W.~B. Croft, ``A transformer-based embedding model for personalized product search,'' in \emph{Proceedings of the 43rd International ACM SIGIR Conference on Research and Development in Information Retrieval}, 2020, pp. 1521--1524.

\bibitem{CroftZAM}
Q.~Ai, D.~N. Hill, S.~Vishwanathan, and W.~B. Croft, ``A zero attention model for personalized product search,'' in \emph{Proceedings of the 28th ACM International Conference on Information and Knowledge Management}, 2019, pp. 379--388.

\bibitem{jd.com}
H.~Zhang, S.~Wang, K.~Zhang, Z.~Tang, Y.~Jiang, Y.~Xiao, W.~Yan, and W.-Y. Yang, ``Towards personalized and semantic retrieval: An end-to-end solution for e-commerce search via embedding learning,'' in \emph{Proceedings of the 43rd International ACM SIGIR Conference on Research and Development in Information Retrieval}, ser. SIGIR '20.\hskip 1em plus 0.5em minus 0.4em\relax New York, NY, USA: Association for Computing Machinery, 2020, p. 2407–2416.

\bibitem{graphsage}
W.~L. Hamilton, R.~Ying, and J.~Leskovec, ``Inductive representation learning on large graphs,'' in \emph{Proceedings of the 31st International Conference on Neural Information Processing Systems}, ser. NIPS'17.\hskip 1em plus 0.5em minus 0.4em\relax Red Hook, NY, USA: Curran Associates Inc., 2017, p. 1025–1035.

\bibitem{NGCF}
X.~Wang, X.~He, M.~Wang, F.~Feng, and T.-S. Chua, ``Neural graph collaborative filtering,'' in \emph{Proceedings of the 42nd International ACM SIGIR Conference on Research and Development in Information Retrieval}, ser. SIGIR'19.\hskip 1em plus 0.5em minus 0.4em\relax New York, NY, USA: Association for Computing Machinery, 2019, p. 165–174.

\bibitem{gcnamazonlu2021graphbased}
H.~Lu, Y.~Hu, T.~Zhao, T.~Wu, Y.~Song, and B.~Yin, ``Graph-based multilingual product retrieval in e-commerce search,'' in \emph{NAACL 2021}, 2021.

\bibitem{ecomranking}
S.~K. Karmaker~Santu, P.~Sondhi, and C.~Zhai, ``On application of learning to rank for e-commerce search,'' in \emph{Proceedings of the 40th international ACM SIGIR conference on research and development in information retrieval}, 2017, pp. 475--484.

\bibitem{clsm}
Y.~Shen, X.~He, J.~Gao, L.~Deng, and G.~Mesnil, ``A latent semantic model with convolutional-pooling structure for information retrieval,'' in \emph{Proceedings of the 23rd ACM International Conference on Conference on Information and Knowledge Management}, ser. CIKM '14.\hskip 1em plus 0.5em minus 0.4em\relax New York, NY, USA: Association for Computing Machinery, 2014, p. 101–110.

\bibitem{matchpyramid}
L.~Pang, Y.~Lan, J.~Guo, J.~Xu, S.~Wan, and X.~Cheng, ``Text matching as image recognition,'' in \emph{Proceedings of the Thirtieth AAAI Conference on Artificial Intelligence}, ser. AAAI'16.\hskip 1em plus 0.5em minus 0.4em\relax AAAI Press, 2016, p. 2793–2799.

\bibitem{matchsrnn}
S.~Wan, Y.~Lan, J.~Xu, J.~Guo, L.~Pang, and X.~Cheng, ``Match-srnn: Modeling the recursive matching structure with spatial rnn,'' in \emph{Proceedings of the Twenty-Fifth International Joint Conference on Artificial Intelligence}, ser. IJCAI'16.\hskip 1em plus 0.5em minus 0.4em\relax AAAI Press, 2016, p. 2922–2928.

\bibitem{geps}
Y.~Zhang, D.~Wang, and Y.~Zhang, ``Neural ir meets graph embedding: A ranking model for product search,'' in \emph{The World Wide Web Conference}, 2019, pp. 2390--2400.

\bibitem{zhuang-etal-2021-robustly}
L.~Zhuang, L.~Wayne, S.~Ya, and Z.~Jun, ``\BIBforeignlanguage{English}{A robustly optimized {BERT} pre-training approach with post-training},'' in \emph{\BIBforeignlanguage{English}{Proceedings of the 20th Chinese National Conference on Computational Linguistics}}.\hskip 1em plus 0.5em minus 0.4em\relax Huhhot, China: Chinese Information Processing Society of China, Aug. 2021, pp. 1218--1227.

\bibitem{xlnetjiang-etal-2020-cross}
Z.~Jiang, A.~El-Jaroudi, W.~Hartmann, D.~Karakos, and L.~Zhao, ``\BIBforeignlanguage{English}{Cross-lingual information retrieval with {BERT}},'' in \emph{\BIBforeignlanguage{English}{Proceedings of the workshop on Cross-Language Search and Summarization of Text and Speech (CLSSTS2020)}}.\hskip 1em plus 0.5em minus 0.4em\relax Marseille, France: European Language Resources Association, May 2020, pp. 26--31.

\bibitem{tk}
S.~Hofst{\"a}tter, M.~Zlabinger, and A.~Hanbury, ``Interpretable \& time-budget-constrained contextualization for re-ranking,'' \emph{arXiv preprint arXiv:2002.01854}, 2020.

\bibitem{xml}
W.-C. Chang, D.~Jiang, H.-F. Yu, C.~H. Teo, J.~Zhang, K.~Zhong, K.~Kolluri, Q.~Hu, N.~Shandilya, V.~Ievgrafov \emph{et~al.}, ``Extreme multi-label learning for semantic matching in product search,'' in \emph{Proceedings of the 27th ACM SIGKDD Conference on Knowledge Discovery \& Data Mining}, 2021, pp. 2643--2651.

\bibitem{hnsdenseeret}
J.~Zhan, J.~Mao, Y.~Liu, J.~Guo, M.~Zhang, and S.~Ma, ``Optimizing dense retrieval model training with hard negatives,'' in \emph{Proceedings of the 44th International ACM SIGIR Conference on Research and Development in Information Retrieval}, ser. SIGIR '21.\hskip 1em plus 0.5em minus 0.4em\relax New York, NY, USA: Association for Computing Machinery, 2021, p. 1503–1512.

\bibitem{hnsnce}
W.~Zhang and K.~Stratos, ``Understanding hard negatives in noise contrastive estimation,'' in \emph{North American Chapter of the Association for Computational Linguistics}, 2021.

\bibitem{distilbertsanh2020}
V.~Sanh, L.~Debut, J.~Chaumond, and T.~Wolf, ``Distilbert, a distilled version of bert: smaller, faster, cheaper and lighter,'' \emph{ArXiv}, vol. abs/1910.01108, 2019.

\bibitem{doct5queryCheriton2019FromDT}
D.~R. Cheriton, ``From doc2query to doctttttquery,'' 2019.

\bibitem{HuangFacebook}
J.-T. Huang, A.~Sharma, S.~Sun, L.~Xia, D.~Zhang, P.~Pronin, J.~Padmanabhan, G.~Ottaviano, and L.~Yang, ``Embedding-based retrieval in facebook search,'' in \emph{Proceedings of the 26th ACM SIGKDD International Conference on Knowledge Discovery \& Data Mining}, 2020, pp. 2553--2561.

\bibitem{zhan2021optimizing}
J.~Zhan, J.~Mao, Y.~Liu, J.~Guo, M.~Zhang, and S.~Ma, ``Optimizing dense retrieval model training with hard negatives,'' in \emph{Proceedings of the 44th International ACM SIGIR Conference on Research and Development in Information Retrieval}, 2021, pp. 1503--1512.

\bibitem{skopt}
\BIBentryALTinterwordspacing
``Scikit-optimize: Sequential model-based optimization in python,'' 2023. [Online]. Available: \url{https://scikit-optimize.github.io}
\BIBentrySTDinterwordspacing

\bibitem{faissjohnson2017billionscale}
J.~Johnson, M.~Douze, and H.~J{\'e}gou, ``Billion-scale similarity search with gpus,'' \emph{IEEE Transactions on Big Data}, vol.~7, pp. 535--547, 2017.

\bibitem{hnswmalkov2018efficient}
Y.~A. Malkov and D.~A. Yashunin, ``Efficient and robust approximate nearest neighbor search using hierarchical navigable small world graphs,'' \emph{IEEE Trans. Pattern Anal. Mach. Intell.}, vol.~42, no.~4, p. 824–836, apr 2020.

\bibitem{fastscancachelocalityandre}
F.~André, A.-M. Kermarrec, and N.~Le~Scouarnec, \emph{Cache locality is not enough: High-Performance nearest neighbor search with product quantization fast scan}, 01 2016, pp. 288--299.

\bibitem{viziergolovin2017google}
D.~Golovin, B.~Solnik, S.~Moitra, G.~Kochanski, J.~Karro, and D.~Sculley, ``Google vizier: A service for black-box optimization,'' in \emph{Proceedings of the 23rd ACM SIGKDD international conference on knowledge discovery and data mining}, 2017, pp. 1487--1495.

\end{thebibliography}

\end{document}